\documentclass{ptapap}

\author{D.\ Baade}[ESO-DE]
\author{Th.\ Rivinius}[ESO-CL]
\author{A.\ Pigulski}[Wroclaw]
\author{A.\ Carciofi}[SaoP]
\author{G.\ Handler}[Warsaw]
\author{R.\ Kuschnig}[Wien, Graz]
\author{Ch.\ Martayan}[ESO-CL]
\author{A.\ Mehner}[ESO-CL]
\author{A.F.J.\ Moffat}[Montreal]
\author{H.\ Pablo}[Montreal]
\author{A.\ Popowicz}[Gliwice]
\author{S.M.\ Rucinski}[Toronto]
\author{G.A.\ Wade}[Kingston]
\author{W.W.\ Weiss}[Wien]
\author{K.\ Zwintz}[Inns]
\affil[ESO-DE]{European Organisation for Astronomical Research in the 
Southern Hemisphere, Karl-Schwarzschild-Str. 2, 85748 Garching b.\ M\"unchen,
Germany}
\affil[ESO-CL]{European Organisation for Astronomical Research in the 
Southern Hemisphere, Casilla 19001, Santiago 19, Chile}
\affil[Wroclaw]{Astronomical Institute, Wroc{\l}aw University, Kopernika 11, 
51-622 Wroc{\l}aw, Poland}
\affil[SaoP]{Instituto de Astronomia, Geof{\' i}sica e Ci{\^ e}ncias Atmosf{\' e}ricas,
Universidade de S{\~ a}o Paulo, Rua do Mat{\~ a}o 1226, 
Cidade Universit{\' a}ria, 05508-900 S{\~ a}o Paulo, SP, Brazil}
\affil[Montreal]{D{\' e}partement de physique and Centre de Recherche en Astrophysique du 
Qu{\' e}bec (CRAQ), Universit{\' e} de Montr{\' e}al, C.P. 6128, 
Succ.\,Centre-Ville, Montr{\' e}al, Qu{\' e}bec, H3C 3J7, Canada}
\affil[Wien]{University of Vienna, Institute for Astrophysics, Tuerkenschanzstrasse 17, 1180 Vienna, Austria}
\affil[Graz]{Graz University of Technology, Institute of Communication Networks
and Satellite Communications, Inffeldgasse 12, 8010 Graz, Austria}
\affil[Toronto]{Department of Astronomy \& Astrophysics, University of Toronto, 
50 St.\,George St, Toronto, Ontario, M5S 3H4, Canada}
\affil[Kingston]{Department of Physics, Royal Military College of Canada, PO
Box 17000, Stn Forces, Kingston, Ontario K7K 7B4, Canada}
\affil[Gliwice]{Institute of Automatic Control, Silesian University of Technology, 
Gliwice, Poland}
\affil[Warsaw]{Nicolaus Copernicus Astronomical Center, ul.\,Bartycka 18, 00-716 
Warsaw, Poland}
\affil[Inns]{Universit\"at Innsbruck, Institute for Astro- and Particle Physics, \\
  Technikerstrasse 25/8, A-6020 Innsbruck, Austria}

\title{Pulsations and outbursts in Be stars:  Small differences -- 
big impacts\footnote{Based on data collected by the BRITE Constellation
satellite mission, designed, built, launched, operated and supported by
the Austrian Research Promotion Agency (FFG), the University of Vienna,
the Technical University of Graz, the Canadian Space Agency (CSA), the
University of Toronto Institute for Aerospace Studies (UTIAS), the
Foundation for Polish Science \& Technology (FNiTP MNiSW), and National
Science Centre (NCN).}}

\begin{document}

\maketitle

\begin{abstract}

{\bf Abstract:} New high-cadence observations with BRITE covering 
many months confirm that coupled pairs of
nonradial pulsation modes are widespread among early-type Be stars.
With the difference frequency between the parental variations they
may form a roughly sinusoidal variability or the amplitude may
cyclicly vary.  A first - amplified - beat pattern is also found.  In
all three cases the amplitudes of difference frequencies can exceed
the amplitude sum of the base frequencies, and modulations of the
star-to-circumstellar-disk mass-transfer rate may be associated with
these slow variations.  This suggests more strongly than any earlier
observations that significant dissipation of pulsational energy in the
atmosphere may be a cause of mass ejections from Be stars.  A unifying
interpretative concept is presented.

\end{abstract}

\section{Introduction}

In a
multi-parameter space, Be stars occupy a volume that is disjoint with all
other types of hot stars, regardless of their evolutionary stage:
\begin{list}{$\bullet$}{\topsep=0mm\parsep=0mm\itemsep=0mm} 
\item
  Be stars show the largest equatorial velocities of all non-degenerate
  stars.
\item
  Be stars possess self-ejected Keplerian disks, which may develop as well as 
  completely disperse on timescales of months to years.
\item
  In UV resonance lines, Be stars exhibit classical wind profiles.  These
  are probably mainly due to radiative ablation of the disk.  
\item
  Companion stars with periods less than a month are rare.  
\item
  At least on large scales, Be stars are the least magnetic group of
  hot stars.  
\end{list} 
For a broader initiation to the Be phenomenon see \citet{2013A&ARv..21...69R}.

In many B stars, rapid rotation and nonradial pulsation seem to be
necessary conditions for this so-called Be phenomenon to occur, as the
rest of this paper will demonstrate.  The rapid rotation may be native
\citep{2007A&A...462..683M} or acquired along the evolutionary path of
a binary \citep{2005ApJS..161..118M}: 
\begin{list}{$\bullet$}{\topsep=0mm\parsep=0mm\itemsep=0mm}
\item
  In a few Be stars, a hot sub-dwarf O-type (sdO) companion has been
  detected by its
  He-ionizing effect on the disk \citep[e.g.,][]{2016ApJ...828...47P}.
  The progenitor of the sdO has dumped
  mass and angular momentum on the present Be primary.  
\item
  Some Be stars have high space velocities
  \citep{2001ApJ...555..364B} that could be the result of 
  the disruption of a binary by either a close encounter in a dense
  environment or the supernova explosion of a former massive companion.
\item
  In Be X-ray binaries, a supernova product is still present in the
  form of a neutron star (evidence for black holes is weak)
  \citep{2015A&ARv..23....2W}.
  \end{list} 
Even the sum of these three processes would probably 
  not account for Be-star frequencies
  reaching $>30$\% at spectral type B1
  \citep{1997A&A...318..443Z}. Most Be-star disks, if any, are not
  accretion disks
  evidencing ongoing mass transfer in a binary \citep[but see
  also][]{2002A&A...396..937H}, and viscous processes and radiative
  ablation destroy Be disks in a few years \citep{2014ApJ...785...12H,
  2016MNRAS.458.2323K} so that they do not trace earlier
  evolutionary stages.  In classical Be stars, the disk matter is decreted
  by the central B star.  Viscosity
  enables tossed-up matter to exchange angular momentum so that
  $\sim$1\% can attain Keplerian velocities while the bulk of the
  ejecta falls back to the star \citep{2014ApJ...785...12H}.

Earlier spectroscopy \citep[e.g.,][]{1988A&A...198..211B} found rapid
outbursts in Be stars.  They even repeat cyclically
\citep{1998A&A...333..125R}, when two NRP modes are temporarily in
phase and co-add their amplitudes. BRITE \citep{2014PASP..126..573W}
has enabled the most detailed description to date of these processes
\citep{2016arXiv161002200B, 2016A&A...588A..56B}, which is expanded
below.  More holistic analyses using wavelet transforms of CoRoT and
Kepler space photometry were performed by \citet[][and in
prep.]{2016arXiv160802872R}.  But it still remains unclear whether all
outbursts of a given Be star are due to the same mechanism and whether
pulsations play a decisive role in all Be stars.

\section{BRITE observations of selected Be stars}

Pre-BRITE observations indicating a link between NRP and mass loss
were often attributed to multi-mode beat processes.
\citet{1998A&A...333..125R} inferred this from the modulation of the
strength of the H$\alpha$ emission of $\mu$ Cen with frequency
differences among a few NRP modes.  In photometry, outbursts are
typically diagnosed from changes in the mean brightness.  They
probably result from varying amounts of circumstellar matter, which
reprocesses the stellar flux it receives and partly redirects it to,
or away from, the observer, depending on viewing angle
\citep{2014ApJ...785...12H}.  In the additive case, it is kind of an
active light echo.  Space photometry contributed the important fact
that power spectra can look substantially different during outbursts
than during quiescence.  Some individual NRP amplitudes decrease, many
increase, and especially quasi-dense groups of not strictly constant
frequencies develop or grow \citep{2009A&A...506...95H,
2016arXiv160802872R}.  Accordingly, the picture was generalized to the
notion that a large number of NRP modes may, in a collective beat
process, energize outbursts \citep[e.g.,][]{2009A&A...506...95H,
2015MNRAS.450.3015K}.

Unexpectedly, BRITE-Constellation has not yet found beat phenomena but
only genuine NRP mode couplings which present themselves as a roughly
sinusoidal large-amplitude variability with a frequency close to the
difference (called $\Delta$ frequency below) between the two parent
frequencies \citep{2016arXiv161002200B, 2016A&A...588A..56B}.  There
are indications that mass-loss rates can be modulated with $\Delta$
frequencies.

The next two subsections illustrate both types of mode combinations;
the third one describes observations that do not fall into either of
these categories.  They extend the previous overview by
\citet{2016arXiv161002200B}.  `Blue' and `red' denote observations
with blue- or red-sensitive BRITE satellites
\citep{2014PASP..126..573W}.

\subsection{Large-amplitude $\Delta$ frequencies} 
\label{delta}

\begin{figure}
\begin{center}
\includegraphics[width=10.4cm]{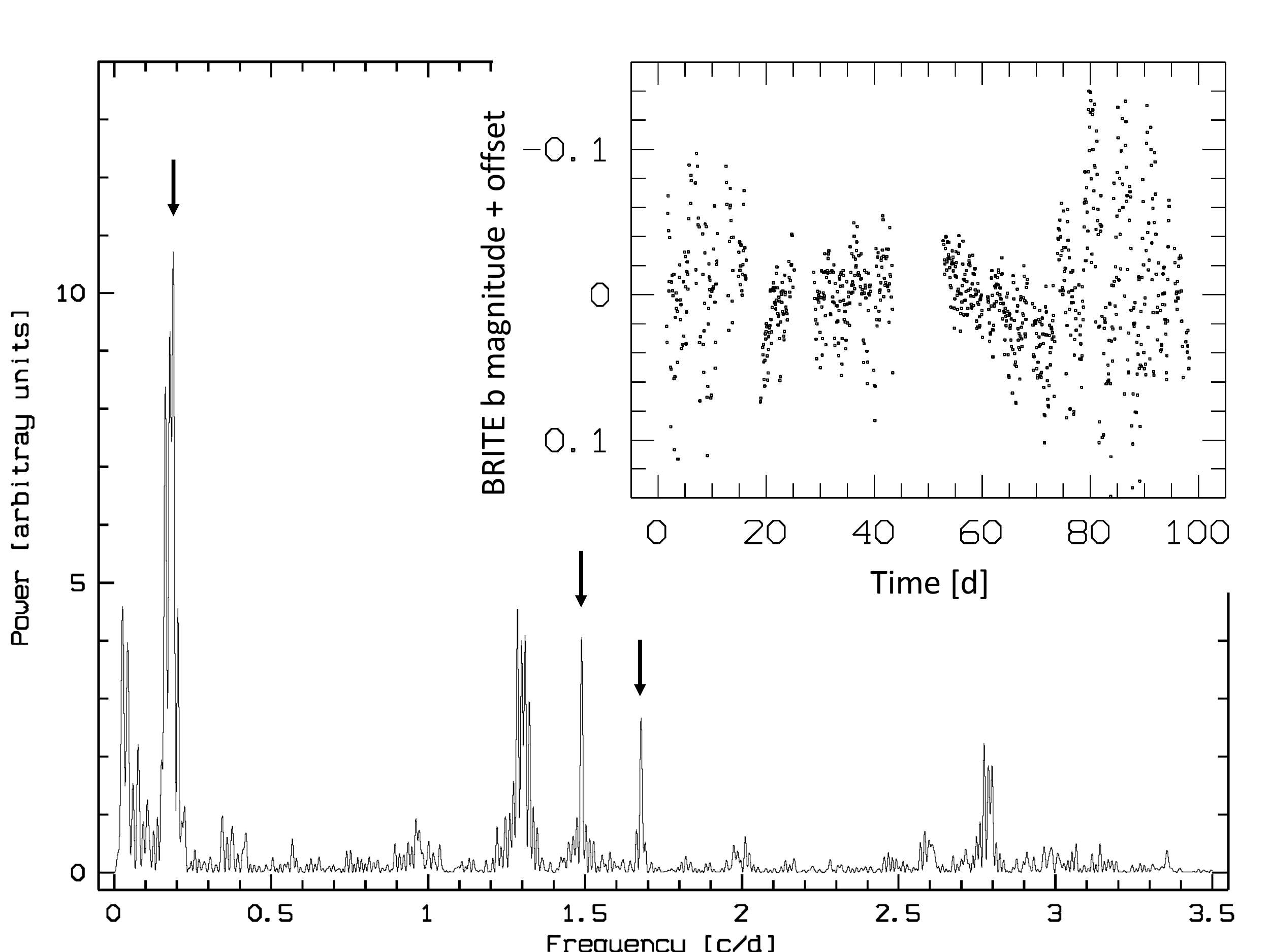}
\end{center}
\caption{Power spectrum of the BRITE blue-channel observations 
of 25 $\psi^1$ Ori in 2014.  Arrows mark the $\Delta$ 
frequency and its parent frequencies.  
The insert shows the lightcurve.}
\label{25Ori}
\end{figure}

The initial prototypes of this variability were $\eta$ Cen
\citep{2016A&A...588A..56B} and 28 Cyg (Baade et al., in prep.).  New
examples include:
\begin{list}{$\bullet$}{\topsep=0mm\parsep=0mm\itemsep=0mm}
  \item
{\bf 10 CMa} (B2, m$_{\rm V}$\,=\,5.2): In 2015, the (blue) amplitude, 13.7\,mmag, of the
    $\Delta$ frequency at 0.0140\,c/d was slightly larger than the
    amplitude sum of its parent frequencies: 1.3363\,c/d (7.7\,mmag) and
    1.3501\,c/d (5.1\,mmag).  There were also strong frequency groups at 0.55\,c/d -
    0.75\,c/d and around 1\,c/d.
  \item {\bf 27 CMa} (B3, m$_{\rm V}$\,=\,4.7): In 2015, a (red) $\Delta$
  frequency, 0.0560\,c/d (2.0 \,mmag), appeared as the difference
  between 2.6825\,c/d (2.3\,mmag) and 2.6257\,c/d (2.4 mmag).  
  There were further frequencies at 1.2678\,c/d and
   1.3573\,c/d with comparable
    (semi-) amplitudes, namely 2.1\,mmag and 1.8\,mmag.  The largest amplitude, 4.4\,mmag, was associated with a fairly isolated peak at 0.7920\,c/d.

   \item {\bf 25 $\psi^1$ Ori} (B1, m$_{\rm V}$\,=\,5.0): This star
   abounds in frequency groups so that plenty of $\Delta$ frequencies
   can be construed \citep[cf.][]{2016arXiv161002200B}. The two 
relatively most
   isolated peaks in the 2015 blue power
   spectrum appeared at 1.4889\,c/d and 1.6784\,c/d with large amplitudes:
   12.7\,mmag and 10.4\,mmag. In other Be stars, amplitudes above
   $\sim$ 10\,mmag often identify the strongest
   in a group and/or are {\v S}tefl frequencies
   \citep{2016A&A...588A..56B, 2016arXiv160802872R}, both of which
   have at least partially strong and variable circumstellar roots.
   Clear frequency groups existed around 0.2\,c/d, 1.3\,c/d, and 2.8\,c/d.

   The strongest power peak in the first group was at 0.1886\,c/d
   (20.3\,mmag), which is close to 0.1895\,c/d\,=\,1.6784\,c/d -
   1.4889\,c/d and, therefore, a $\Delta$ frequency.
   Fig.\,\ref{25Ori} reveals in full detail what dramatic effects can
   be
   associated with $\Delta$ frequencies.  The
   overall structure of the light curve is governed by two series of outbursts,
   one each at the beginning and end of its time coverage.  On shorter
   timescales, the outbursts imprint a conspicuous structure with the 
   $\Delta$ frequency, 0.19\,c/d.  But
   there is no beat pattern of the parent frequencies, 1.49\,c/d and
   1.68\,c/d.  Neither does the $\Delta$ frequency
   appear as a steady sinusoidal modulation as in $\eta$ Cen
   \citep{2016A&A...588A..56B} and 28 Cyg (Baade et al., in prep.).
   Instead, the \underline{amplitude} of the $\Delta$
   frequency is strongly modulated.  
   Unfortunately, the time span of the observations is not
   long enough to reveal whether the series of outbursts repeat every
   three months.

The light curve contains an additional special detail.
During the outburst series, there are not just brightenings, which
probably are circumstellar light echoes, but also fadings.  They occur
in both blue and red data confirming they are real variations.  The
nature of the effects of ejected matter, brightening or dimming,
depends on aspect angle \citep{2014ApJ...785...12H}.  In a given star,
this angle is constant, and so the observations contain constraints on
the width in stellar latitude of the mass-exchange process between
star and disk.
   \end{list}

In $\gamma$ Cas (B0.5, m$_{\rm V}$\,=\,2.4) two parent variations
(2.1950\,c/d, 0.9\,mmag and 2.4796\,c/d, 2.7\,mmag), of which the
weaker one may be a member of a frequency group, gave in 2015 rise to
a $\Delta$ frequency at 0.27\,c/d and low amplitude (1.1\,mmag).  A
second major variability (2.2\,mmag) was at 0.9728\,c/d.

\begin{figure}
  \centering
  \begin{minipage}{0.48\textwidth}
    \includegraphics[width=\textwidth]{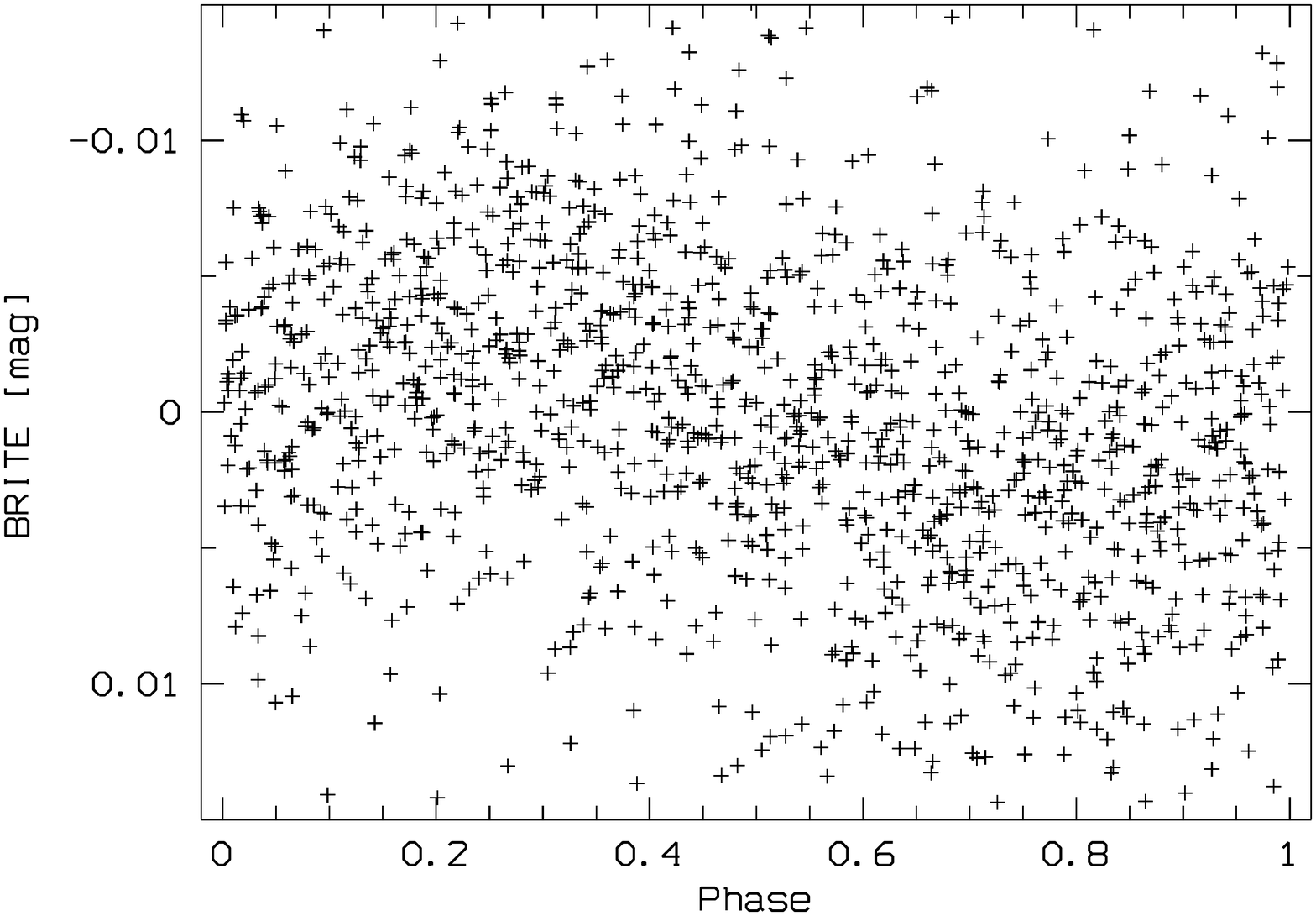}
    \caption{BRITE photometry of $\gamma$ Cas \newline folded with $f$\,=\, 2.479\,c/d.}
    \label{gamCasfast}
  \end{minipage}
  \quad
  \begin{minipage}{0.48\textwidth}
    \includegraphics[width=\textwidth]{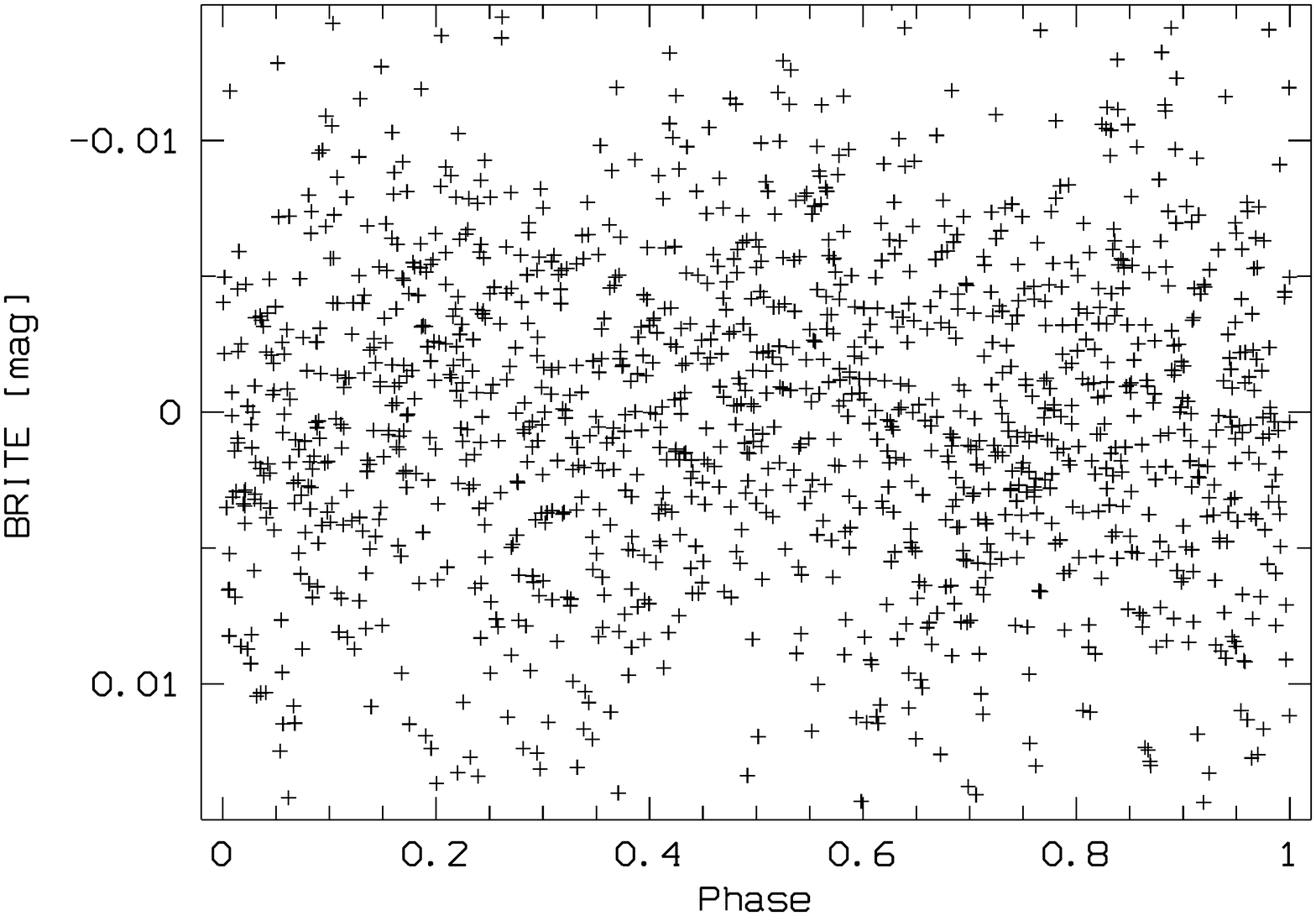}
    \caption{Ditto except for $f$\,=\,0.8225\,c/d.~~~~~~~~~}
\label{gamCasslow}
  \end{minipage}
\end{figure}

Among these inconspicuous frequencies, 2.4796\,c/d is the most
interesting (Fig.\ \ref{gamCasfast}).  It is close to three times 
0.8225\,c/d (0-3\,mmag) which \citep{2012ApJ...760...10H} found in 15
seasons (1997-2011) of single-site ground-based photometry.  Our
analysis of 7 years (2002-2011) of observations with SMEI
\citep{2004SoPh..225..177J} confirms its reality.
\citet{2016AdSpR..58..782S} use this frequency as an anchor quantity
of their magnetically-controlled rotational-modulation model which
interprets it as the rotation period of the primary in the 200-d
binary.  This 0.8225-c/d frequency is not seen in the BRITE
observations (Fig.\ \ref{gamCasslow}).  A three times faster rotation
would stretch the model of \citet{2016AdSpR..58..782S} beyond its
limits because its authors state that even their lower frequency would
imply 1.15 $\pm$ 0.15 fractional critical rotation.  They propose
$\gamma$ Cas as the prototype of hard X-ray emitting Be stars.  The
BRITE power spectrum appears rather normal but the apparent frequency
tripling is quite unusual.

\subsection{Beat phenomena}
\label{beat}

The spectroscopic frequencies and difference frequencies
\citep{1998A&A...333..125R} of $\mu$ Cen (B2, m$_{\rm V}$\,=\,3.4)
were not detected with BRITE \citep{2016A&A...588A..56B}.  The
large-amplitude ($\sim$250\,mmag peak to peak in the red BRITE
passband) variability is due to a light echo, which varies with the
amount of matter present in the inner disk.  Conceivably, these light
echoes are not good clocks so that the stellar frequencies are not
(well) reproduced.  $\kappa$ CMa (m$_{\rm V}$\,=\,3.7) not only has a
similar spectral type (B1.5 vs.\ B2) but also a comparable
$v$\,sin\,$i$ (220\,km/s vs.\ 155\,km/s) and is probably viewed at a
similar, perhaps slightly higher, inclination angle.
\citet{2003A&A...411..229R} found two spectroscopic frequencies,
0.548\,c/d and 0.617\,c/d and also discuss discrepant photometric
frequencies; the line-profile variability associated with the lower
frequency is different from the typical $\ell$\,=\,$+m$\,=\,2 variety.
They are not visible in the BRITE red photometry from 2015, which is
dominated by outbursts that repeat cyclicly with about 0.0557\,c/d.
All three frequencies deserve further observational attention because
the large range of $\sim$150\,mmag suggests that the low frequency is
due to a light echo.  The proximity to a $\Delta$-frequency relation
may be just spurious, or one or more of the frequency values are
incorrect.  However, if the deviation from a $\Delta$-frequency
relation contains a component from the orbital phase velocity of an
inhomogeneous or non-circular circumstellar mass distribution, it
could carry important diagnostics of the star-disk interaction region.

\begin{figure} \begin{center}
\includegraphics[width=10.4cm]{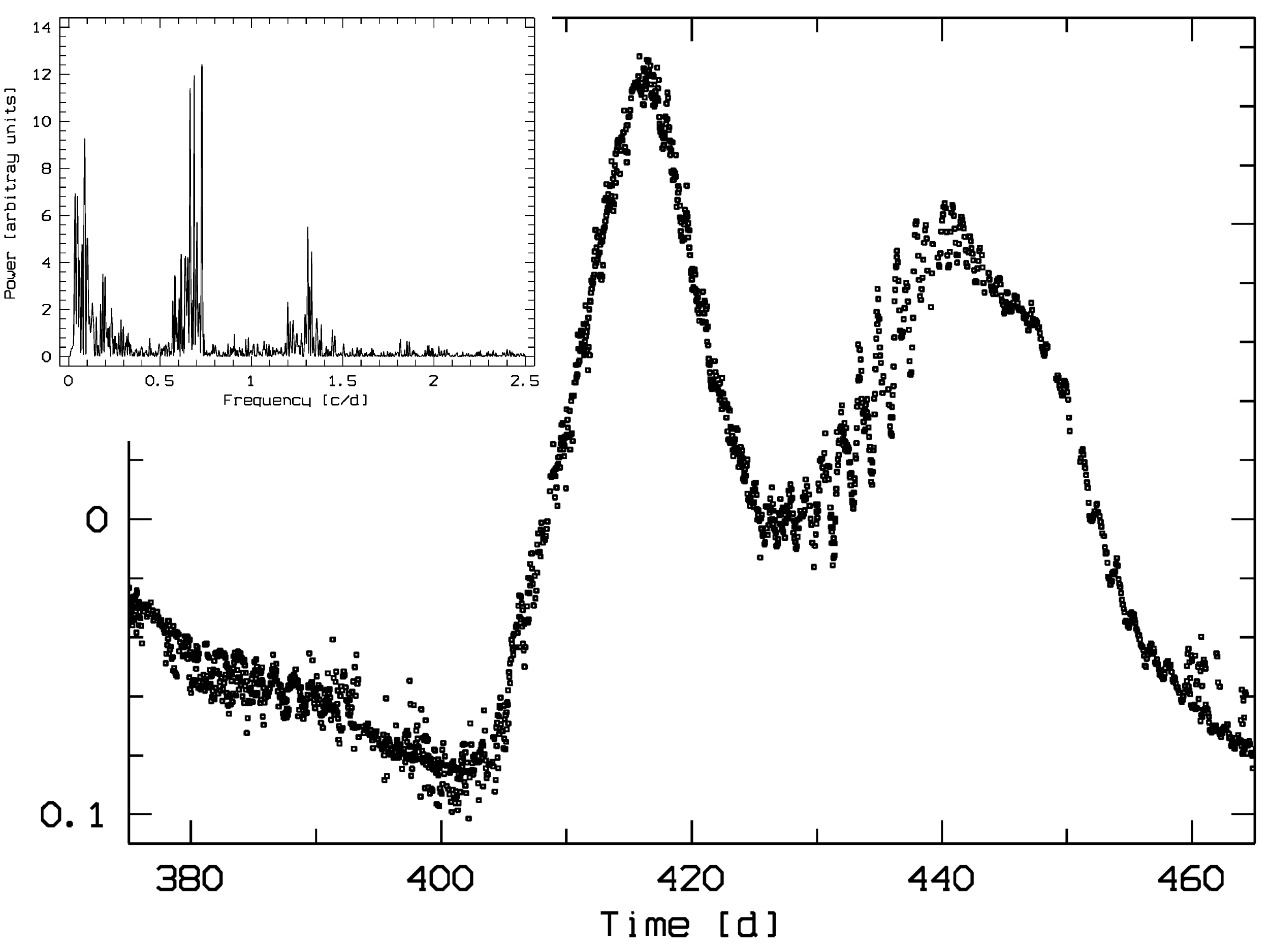} 
\end{center} 
\caption{Red
light curve of 28 CMa obtained with BRITE in 2015.  Note the various
phases with enhanced activity, very similar to beat patterns and most
prominent from 430\,d to 440\,d where the spacing of the wiggles is
consistent with frequencies in the group near 0.7\,c/d.  The insert
displays the power spectrum of the data corrected for the very
high-amplitude, slow light-echo variability.}  
\label{28CMa}
\end{figure}

In all the many BRITE Be-star light curves, genuine beat
patterns are a rare sight although all stars are multi-frequency
variables.  They may be veiled by light echoes and atmospheric
amplitude amplification processes.  Moreover, members of dense
frequency groups, which are probably largely circumstellar
\citep{2016A&A...588A..56B, 2016arXiv160802872R}, are often not
strictly phase coherent and therefore cannot form beat patterns.
But the paucity of beat patterns is still remarkable.

28 $\omega$ CMa (B2, m$_{\rm V}$\,=\,3.8) is one of the first Be stars
in which nonradial pulsation was suspected
\citep{1982A&A...105...65B}.  The initial frequency of 0.73\,c/d was
approximately confirmed by various other spectroscopic studies
\citep[e.g.,][]{1999MNRAS.305..505S}. In the blue BRITE power spectrum
(data from 2015) only a low-significance peak at 0.7273\,c/d is found.
This appears not surprising because in pole-on stars like 28 $\omega$
CMa ($v$ sin $i$\,=\,80\,km/s), the latitudinal component of the
velocity field of quadrupole NRP modes is maximally visible
\citep{2003A&A...411..229R} whereas the photometric signature may well
be azimuthally averaged out to near-invisibility.  However, after
subtraction of the enormous $\sim$150/200-mmag (blue/red) peak-to-peak
variability, which is a light echo as in $\mu$ Cen
\citep{2016A&A...588A..56B}, $\kappa$ CMa (Sect.\,\ref{delta}), and
many others, it becomes evident that 0.73\,c/d is amidst a frequency
group extending from 0.55\,c/d to 0.8\,c/d.  This explains the
insuccessful efforts to improve the spectroscopic frequency.  The
group also includes this star's {\v S}tefl frequency at 0.68\,c/d
\citep{1999MNRAS.305..505S}. The mean (semi-)amplitudes of the highest
peaks reach $\sim$2.5\,mmag.  A second frequency group exists between
1.15\,c/d and 1.45\,c/d.

In Fig.\ \ref{28CMa}, the power spectrum after removal of the slow
large-scale variability and part of the un-cleaned light curve are
presented.  The most important detail lies in the series of wiggles
identified in the caption to the figure.  The wiggle spacing is
consistent with frequency values in the 0.55\,c/d - 0.8\,c/d group.
Their peak-to-peak range, $\sim$50\,mmag, is in the light-echo domain
and an order of magnitude larger than the amplitudes in the frequency
group.  This large amplification supersedes the expected classical
beat pattern. Since these tremolos are visible most of the time, it is
not currently possible to conclude whether they can be the cause of
the slow large-amplitude variability, i.e.\ major ejections of matter.
Special interest exists in finding out whether the series of outbursts
repeat every $\sim$75 days as seems possible from the present data.
1/75\,d\,=\,0.013\,c/d could, then, be called a $\Delta$ frequency.

\subsection{Orphan frequencies and other variabilities} 

Not all low-frequency variabilities can be traced back to a difference
between two higher (NRP) frequencies.  Some of these `orphan
frequencies' are of the order of 2/T, where T is the total time span
of the observations, but closely spaced frequency peaks should still
be resolvable.  The existence of $ \Delta$ frequencies with similar
properties suggests that there is a good chance of many of the orphan
frequencies also being real. Independent support may be derived from
the report by \citet{2016arXiv160908449L} of even lower frequencies
(down to $\sim$ 0.005\,c/d) in ground-based long-term photometry of 610
Be stars. BRITE examples include:

\begin{list}{$\bullet$}{\itemsep=0mm\parsep=0mm\topsep=0mm}

\item {\bf $\omega$ Ori} (B3, m$_{\rm V}$\,=\,4.6) exhibited slow
(red) variability with 0.0275\,c/d (9.4\,mmag) in 2014 and 0.0416\,c/d
(7.6 mmag) in 2015.  In 2014, the main (red) rapid variability was at
1.0519\,c/d (9.9\,mmag) but in 2015 there were instead two similarly
strong variations at 0.9872\,c/d (6.9\,mmag) and 1.0719\,c/d
(6.5\,mmag).  All three higher frequencies are embedded in a frequency
group that is broad enough to easily construct from its ranks the low
frequencies as $\Delta$ frequencies.

\item {\bf FW CMa} (B2, m$_{\rm V}$\,=\,5.2): The red light curve
(from 2015) of this extreme pole-on star \citep[40
km/s,][]{2003A&A...411..229R} is not too badly approximated by a
single frequency, 0.0506\,c/d.  Maxima and minima are fairly peaked
(as opposed to sinusoidal), and small changes in shape and
considerable amplitude variations give rise to the suspicion that the
true nature of the variability is a beat process.  But the power
spectrum does not reveal any candidate parent frequencies; the
spectroscopic frequency of 1.192\,c/d \citep{2003A&A...411..229R} is
not detected.  The amplitudes grew over four months from
$\sim$10\,mmag to $\sim$ 30\,mmag.  Overall, this light curve of FW
CMa can be described as being very similar to that of 25 $\psi^1$ Ori
except for a stretch in time by a factor of a few.

\end{list}

\noindent
A special, so far singular, case is 60 Cyg (B1, m$_{\rm
V}$\,=\,5.4).  In 2015, it featured two (red) frequency pairs: (i)
1.8672\,c/d (2.4\,mmag) and 1.8867\,c/d (3.3\,mmag) as well as (ii)
3.7538\,c/d (2.6\,mmag) and 3.7726\,c/d (2.4\,mmag).  In both pairs
the frequencies differed by 0.19\,c/d.  Moreover, the higher frequencies
in the two pairs are harmonics.  But there was no obvious variability
with 0.19\,c/d.  At 3.3336\,c/d, 60 Cyg had the strongest
(6.4\,mmag) variability above 2.5\,c/d so far seen with BRITE.

\section{Summary, conclusions, and outlook}

Observations with BRITE-Constellation have established that in Be
stars pairs of NRP modes can couple to form a sinusoidal variability
with their difference ($\Delta$) frequency or a beat pattern.  Both
may reach amplitudes above the amplitude sum of the parent modes,
which signals (as light echoes), maybe enables, mass-loss outbursts.
The coupling seems highly selective, and most combinations of NRP
modes have no visible effect (this conclusion assumes broadly 
correct distinction between stellar NRP modes and circumstellar
frequencies, see below).  The sensitivity of BRITE is not sufficient
to investigate the behaviour of frequency groups because the
frequencies are time dependent. They may harbour a similar mechanism
involving the collective power of many elementary variabilities.
While the evidence for outbursts due to coupled NRP modes seems
overwhelming, such evidence is not found in similarly many other Be
stars observed by BRITE.  Explanations include:
\begin{list}{$\bullet$}{\itemsep=0mm\parsep=0mm\topsep=0mm} 
\item
  The stars are inactive for extended periods of time.
\item
  The aspect-angle-dependent amplitudes of NRP modes
  are too low.
\item
The stellar variability is masked by circumstellar light echoes.  
\end{list} 
All stars discussed in this paper are of spectral
  type B3 or earlier.  This introduces a strong bias towards higher
  activity \citep{2016arXiv161002200B, 2016arXiv160908449L}.

A larger sample would permit searches for trends with aspect angle
($v$\,sin\,$i$) and for differences between stellar and circumstellar
variabilities.  An estimator of the latter may be the environment in
the power spectrum.  If it consists of a rich group, the phase
coherence is often lower \citep{2016A&A...588A..56B,
2016arXiv160802872R}, which might be more readily reconciled with
circumstellar conditions.  Amplitudes above $\sim$10 mmag may also
more commonly have a circumstellar component (see below).  If
quadrupole modes are also photometrically dominant, the 
nearly perfect pole-on star FW CMa demonstrates that the
large-amplitude variability is circumstellar.  There is no obvious
wavelength dependency of the amplitudes but colour variations may be
different for stellar (bluer?) and circumstellar (redder?) variations.

Multi-season observations would gain in scope towards repetition
frequencies of major outbursts, which may be as low 0.0003\,c/d or
less \citep[e.g.,][]{2015arXiv150608902G}, and also enable searches
for the origins of orphan frequencies.  Other goals include the
comparison of the angular 4$\pi$ structure of $\Delta$ variations to
that of their parent NRP modes and the localization of the amplitude
amplification process.

The central task is the development, if possible, of a unifying
scheme.  It could look as follows: Via some unknown selection
mechanism, two NRP modes couple to form a $\Delta$ frequency. Because
of the increased combined amplitude (and with much rotational
support), they cause cyclicly repeating mass-loss outbursts.  Large
amplitudes of $\Delta$ frequencies arise as circumstellar light
echoes.  They trace the combined and integrated effects in the inner
disk of mass injection and viscous and radiative gas dispersal.  This
can be different from the amount of mass instantaneously ejected. 

In $\eta$ Cen and 28 Cyg, the amount of inner-disk matter or its
excitation varies sinusoidally with a $\Delta$ frequency of constant
amplitude.  If it is excitation, it could be a lighthouse effect
caused by the pulsationally varying and propagating photospheric
radiation field.  It was expected \citep{1986PASP...98...35P} for
individual low-order modes.  To date, there is no observational
confirmation.  But perhaps the power associated with $\Delta$
frequencies is more effective.  In 25 $\psi^1$ Ori and FW CMa, this
light modulation exhibits a time-dependent amplitude and is akin to a
beat pattern.  Only 28 CMa exhibits bona-fide beat patterns of
frequencies in the NRP range (which is slightly surprising as the
frequencies in the groups are not really phase coherent). Within the
light-echo paradigm this suggests that effects of individual pulsation
cycles are seen. In the absence of other evidence, one can only
speculate whether in the other stars there may be a high-frequency
cut-off somewhere.

Particularly interesting is the amplitude amplification of the higher
frequencies.  This could also be the base mechanism of the $\Delta$
frequency amplitude variability in 25 $\psi^1$ Ori and FW CMa: The
beat amplitude is not enhanced as such but merely as a consequence of
growing amplitudes of the base variabilities.  The reason for the
variation in amplitude of the higher frequencies may be in the disk or
the star or both.  Finally, if several $\Delta$ frequencies coexist,
as perhaps in $\mu$ Cen and 28 CMa, the superimposed light echoes can
be impossible to unravel from `just' half a year worth of data,
especially if not fully periodic.

The extension in timescale to a decade may require correspondingly low
$\Delta$ frequencies, which would hardly be directly detectable.
However, if there were two (or more) outburst patterns which can be
distinguished as patterns, extremely low super$\Delta$ frequencies
could be indirectly derived. The duration of the associated outbursts
would, then, depend on the length of the phase interval, during which
high mass loss can be sustained by the involved pairs of NRP modes.

The two-coupled-NRP-modes paradigm is a useful approximative
description of the observations.  If the frequencies are not constant
and embedded in a dense group, the overall process may be highly
stochastic if not chaotic.  This could include rogue-wave-like events. 
Where present, the semi-regular repetition of outbursts would argue
against excessive randomness.  But in many Be stars observed at
sufficient spectral resolution and S/N, higher-order line-profile
variability has been detected \citep{1983ApJ...275..661V}, which
provides an additional activity floor.

\acknowledgements{\footnotesize AFJM is grateful for financial aid from
CSA, NSERC and FQRNT (Quebec).  OK, RK and WW are grateful for funding
via the Austrian Space Application Programme (ASAP) of the Austrian
Research Promotion Agency (FFG) and BMVIT; and WW for support of the
University of Vienna (IS 538001, IP 538007). KZ acknowledges support
by the Austrian Fonds zur Foerderung der wissenschaftlichen Forschung
(FWF, project V431-NBL). SMR acknowledges support from NSERC
of Canada. The Polish co-authors gratefully acknowledge
support from the BRITE PMN grant 2011/01/M/ST9/05914; GH additionally
thanks the NCN grant 2015/18/A/ST9/00578. JMM and GAW acknowledge
Discovery Grant support from NSERC. AP acknowledges the NCN grant No.
2011/03/B/ST9/02667. APo was supported by the Polish National Science
Center, grant No.\ 2013/11/N/ST6/03051.  }

\bibliographystyle{ptapap}


\end{document}